\documentclass[aps,prb,showpacs,balancelastpage,amssymb,groupedaddress]{revtex4}

\begin{document}

\title{PUMPING CURRENT IN A QUANTUM DOT  BY AN OSCILLATING MAGNETIC FIELD }

\author{ZHAO YANG ZENG}

\address{ Department of of Physics, Jiangxi Normal
University, Nanchang 330027, China\\}

\author{HONG LI}

\address{$^1$ Department of Physics, Huaibei coal  industry teachers college, Huaibei,235000, China\\
$^2$ Department of of Physics, Jiangxi Normal University, Nanchang
330027, China}

\author{ BAOWEN LI}
\address{ $^1$ Department of Physics and Centre for Computational Science and Engineering, National University of Singapore, 117542, Singapore\\
  $^2$Laboratory of Modern Acoustics and Institute of Acoustics, Nanjing University, Nanjing 210093, PR China\\
 $^3$ NUS Graduate School for Integrative Sciences and
Engineering, Singapore 117597,  Singapore \\}

\begin{abstract}

We investigate spin and charge current through a quantum dot
pumped by a time-varying magnetic field.    Using the density
matrix method, quantum rate equations for the electronic
occupation numbers in the quantum dot are obtained and solved in
the stationary state limit for a wide set of setup parameters.
Both charge and spin current are  expressed explicitly in terms of
several relevant parameters and analyzed in detail. The results
suggest a way of optimizing experimental setup parameters to
obtain an maximal spin current without the charge current flow.

\end{abstract}


\maketitle

Spintronics is an emerging active research field, which is based
on the effective control of electron spin in addition to its
charge degree of freedom\cite{1}.  One common operational
principle for spin-based devices is  how to generate a spin
current. It is known that  a spin current
 is usually accompanied by a charge current, which generates heat. And this may pose a severe problem in microelectronics as the devices become
 smaller and smaller, since heat destabilizes the operation of the
 devices.  Therefore  proposals for the spintronics devices without creation of charge current flow would be greatly desirable.

 Some spin-based devices have been already proposed\cite{2}. However, one of the most effective way to create a spin current is
 using semiconductor quantum dots\cite{3,4,5,6,7,8}. Martin et al.
 proposed a scheme for electrical detection of electron spin
 resonance(ESR) of an electron trap\cite{5}. This idea intrigues
 subsequent studies on how to achieve a desirable spin current
 through a quantum dot or multiple quantum dots either in the
 strong Coulomb regime\cite{5,7,8} or in the Kondo regime\cite{6}.
The basic idea is first Zeeman splitting the dot level by a
perpendicular constant magnetic field and then pumping electron
from the low-lying spin-up state to the high-lying spin-down state
by an oscillating magnetic field. We notice that special
parameters are chosen to facilitate the studies\cite{5,6,7,8}. It
is the purpose of this work to relax the parameter constraints to
gain a  general physical picture about spin pump effect in a
quantum dot by a time-varying magnetic field.

The quantum dot spin pump consists of a quantum dot coupled to two
electronic leads
 by tunnel barriers. No voltage bias is applied to the
device to allow for a direct current through the quantum dot.  To
realize a spin pump with such a system, one apply a perpendicular
constant magnetic field $B_0 \hat{z}$  and  a lateral time-varying
field $B_{rf}[cos(\omega_{rf}t) \hat{x}+sin(\omega_{rf}t)\hat{y}]$
to the quantum dot. Due to Zeeman effect in the presence of a
constant magnetic field $B_0 \hat{z}$, the dot level will be split
into two spin-dependent levels:
$\varepsilon_d\rightarrow\varepsilon_{\sigma}=\varepsilon_d-\sigma
E_Z(=g\mu_BB_0)/2$, where $g$, $\mu_B$ are the effective $g$
factor and Bohr magneton of the quantum dot. The oscillating
magnetic field serves a machine to pump low-lying spin-up
electrons to the high-lying spin-down level, which is described by
a Hamiltonian $H_{rf}$ given below. We consider the case  that the
Coulomb interaction between electrons inside the dot is strong
enough to prohibit double occupation of the dot. A spin-up
electron tunnels into the quantum dot from either the left or
right lead, and tunnels out of it after being pumped to the
high-lying state. This process persists repeatedly and  a steady
charge and spin current is generated.

The Hamiltonian of this device is
\begin{equation}
\label{total hamiltonian}
 H  =  \sum_\sigma \varepsilon_\sigma d_\sigma^\dagger d_\sigma+\sum_{k\sigma}^{\alpha=l,r}\varepsilon_{\alpha k\sigma}
 c_{\alpha k\sigma}^\dagger c_{\alpha k\sigma}+\sum_{k\sigma}^{\alpha=l,r}t_{\alpha k \sigma}(c_{\alpha k\sigma}^\dagger
 d_\sigma+d_{\sigma}^\dagger c_{\alpha k\sigma})
        +H_{rf}(t),\nonumber
\end{equation}
where $d_\sigma^\dagger(d_\sigma)$ creates(annihilates) an
electron with spin
 $\sigma =\uparrow, \downarrow$ in the quantum dot at the level
$\varepsilon_\sigma$, $c_{\alpha k\sigma}^\dagger( c_{\alpha
k\sigma})$  creates(annihilates) a spin-$\sigma$ electron in the
lead $\alpha=l,r$.  The third term in Eq. (1) describes tunneling
between the dot and the leads, while the last term $H_{rf}$
denotes the pumping mechanism and can be written as $
H_{rf}(t)=\Omega_R(d_{\uparrow}^+d_{\downarrow}e^{i\omega_{rf}t}
+d_{\downarrow}^+d_{\uparrow}e^{-i\omega_{rf}t})/2$, where
$\Omega_R=g\mu_BB_{rf}/2$($\hbar$ is set to be unity throughout
this work) is the Rabi frequency.

 A gate voltage must be applied to the quantum dot and so adjusted
 that the chemical potential of the leads lies between the spin-up
 and spin-down levels of the dot. Initially the system lies in its
 ground state $|G\rangle$ with electrons filling up to the
 chemical potential  in the leads and without electron inside the dot.
 When tunneling is turned on, the wave function of the whole
 system can be written in the following form
\begin{eqnarray}
\label{2.3}
 |\Psi(t)\rangle&=&\Large\{b_0(t)+\sum_{k}^{\alpha=l,r}\big[b_{\alpha k \uparrow}(t)d_{\uparrow}^\dagger
 c_{\alpha k \uparrow}
 +b_{\alpha k \downarrow}(t)d_{\downarrow}^\dagger c_{\alpha k
 \uparrow}\big]
 +\sum_{k,k'}^{\alpha,\beta=l,r}b_{\alpha\beta}(t) c_{\beta k'\downarrow}^\dagger c_{\alpha k\uparrow}+\nonumber\\
&&\sum_{k,k',k''}^{\alpha,\beta,\gamma=l,r}\big[b_{\alpha\beta\gamma\uparrow}(t)d_{\uparrow}^\dagger
c_{\gamma k''\downarrow}^\dagger c_{\alpha k\uparrow}c_{\beta
k'\uparrow}+b_{\alpha\beta\gamma\downarrow}(t)d_{\downarrow}^\dagger
c_{\gamma k''\downarrow}^\dagger c_{\alpha k\uparrow}c_{\beta
k'\uparrow}\big]+\cdots\Large\}|G\rangle,\nonumber
\end{eqnarray}
where $b(t)'s$ denote the probability amplitudes for finding the
system in the corresponding states at time t, and initially  all
the $b(0)'s$ except $b_0(0)$ are zero.

 Now we introduce the reduced density
matrix $\rho_{ij}$ spanned in the Fock space of the quantum dot:
$|0\rangle \rightharpoonup$ empty state,
$|\uparrow\rangle\rightharpoonup$ the spin-up state is occupied ,
$|\downarrow\rangle\rightharpoonup$ the spin-down state is
occupied. The diagonal elements of the density matrix $\rho_{ii}$
give  the probabilities of finding the dot being either empty  or
occupied by a spin-$\sigma$ electron, while the off-diagonal
elements describe coherent superposition of the spin-up and
spin-down states. The density matrix $\rho_{ij}$ can be obtained
by tracing out the degrees of freedom of the leads in the full
density matrix
$\rho_{ij}=\sum_{n_{l\uparrow},n_{l\downarrow};n_{r\uparrow},n_{r\downarrow}}
\rho_{ij}^{(n_{l\uparrow},n_{l\downarrow};n_{r\uparrow},n_{r\downarrow})}$,
where
$\rho_{ij}^{(n_{l\uparrow},n_{l\downarrow};n_{r\uparrow},n_{r\downarrow})}$
represent the probabilities of finding the dot in the state $ij$
with $n_{l\uparrow}$ and $n_{r\uparrow}$ spin-up electrons
tunneling out of the left and right leads, and  $n_{l\downarrow}$
and $n_{r\downarrow}$ spin-down electrons tunneling into the left
and right leads. We find
$\rho_{00}=|b_0(t)|^2+\sum_{k,k'}^{\alpha,\beta=l,r}|b_{\alpha\beta}(t)|^2+\cdots$,
$\rho_{\sigma\sigma}=\sum_{k}^{\alpha=l,r}|b_{\alpha k
\sigma}(t)|^2+\sum_{k,k',k''}^{\alpha,\beta,\gamma=l,r}|b_{\alpha\beta\gamma\sigma}(t)|^2+\cdots$,
$\rho_{\uparrow\downarrow}=\sum_{k}^{\alpha=l,r} b_{\alpha k
\uparrow}(t)b_{\alpha k \downarrow}^*(t)
+\sum_{k,k',k''}^{\alpha,\beta,\gamma=l,r}b_{\alpha\beta\gamma\uparrow}(t)b_{\alpha\beta\gamma\downarrow}^*(t)
+\cdots$.  Solving the schr$\ddot{o}$dinger equation
$i\frac{d|\Psi(t)\rangle}{d t}=H |\Psi(t)\rangle$ results in an
infinite set  of coupled linear differential equations for
$b(t)'s$, which can be finally transformed into an infinite set of
algebraic equations\cite{9,10} after the Laplace transform
$b(E)=\int^\infty_0 dt  b(t)e^{iEt}$.
 Performing inverse Laplace transform and summing up
the relevant terms, we obtain the follow quantum rate equations
for the density matrix $\rho_{ij}$
\begin{eqnarray}
\label{rate eqation}
 \dot{\rho}_{00} & = &-(\Gamma_{l\uparrow}+\Gamma_{r\uparrow})\rho_{00}
 +(\Gamma_{l\downarrow}+\Gamma_{r\downarrow})\rho_{\downarrow\downarrow},\\
 \dot{\rho}_{\uparrow\uparrow} & = &(\Gamma_{l\uparrow}+\Gamma_{r\uparrow})\rho_{00}
 +i\frac{\Omega_R}{2}(e^{i\omega_{rf}t}\sigma_{\uparrow\downarrow}
                -e^{-i\omega_{rf}t}\rho_{\downarrow\uparrow}),\\
 \dot{\rho}_{\downarrow\downarrow} & = &
 -(\Gamma_{l\downarrow}+\Gamma_{r\downarrow})\rho_{\downarrow\downarrow}
 -i\frac{\Omega_R}{2}(e^{i\omega_{rf}t}\rho_{\uparrow\downarrow}
               -e^{-i\omega_{rf}t}\rho_{\downarrow\uparrow}),\\
 \dot{\rho}_{\downarrow\uparrow} & = & (iE_Z-
 \frac{\Gamma_{l\downarrow}+\Gamma_{r\downarrow}}{2}
-\frac {1}{T_\perp})
\rho_{\downarrow\uparrow}+i\frac{\Omega_R}{2}{e^{i\omega_{rf}t}}
                                   (\rho_{\downarrow\downarrow}-\rho_{\uparrow\uparrow}),\\
 \dot{\rho}_{\uparrow\downarrow} & = & (-iE_Z-
 \frac{\Gamma_{l\downarrow}+\Gamma_{r\downarrow}}{2}
-\frac {1}{T_\perp})
\rho_{\uparrow\downarrow}+i\frac{\Omega_R}{2}{e^{-i\omega_{rf}t}}
                                   (\rho_{\uparrow\uparrow}-\rho_{\downarrow\downarrow}),
\end{eqnarray}
where $\Gamma_{\alpha \sigma}=2\pi\sum_{k} |t_{\alpha
k\sigma}|^2\delta(\omega-\varepsilon_{\alpha k\sigma})$ are the
line-width functions characterizing the coupling strength between
the dot and the leads. We have introduced phenomenologically an
additional relaxation term $1/T_\perp$ to describe the transverse
spin relaxation process.

Current $I_{\alpha \sigma}$ is calculated as the evolution rate of
electron number tunneling into or out of the lead $\alpha$:
$I_{\alpha
\sigma}=e\dot{N}_{\alpha\sigma}(t)=\sum_{n_{\alpha\sigma}}^{i}
n_{\alpha\sigma}\dot{\rho}_{ii}^{(n_{l\uparrow},n_{l\downarrow};n_{r\uparrow},n_{r\downarrow})}$.
In the stationary state limit, $\dot{\rho}_{ij}=0$, one finds the
following expressions for the charge current
$I^c_\alpha=I_{\alpha\downarrow}+I_{\alpha\uparrow}$ and the spin
current $I^s_\alpha=I_{\alpha\downarrow}-I_{\alpha\uparrow}$
flowing into the left or right lead

\begin{eqnarray}
  I_l^c& = & I_r^c= e\Omega_R^2(\Gamma_{l\uparrow}\Gamma_{r\downarrow}-
                \Gamma_{l\downarrow}\Gamma_{r\uparrow})
                \Theta(E_Z,\Omega_R, \omega_{rf}, T_\perp, \Gamma_{\alpha\sigma}),\\
 I_{l}^s & = & -e\Omega_R^2(2\Gamma_{l\uparrow}\Gamma_{l\downarrow}+
        \Gamma_{l\uparrow}\Gamma_{r\downarrow}+\Gamma_{l\downarrow}\Gamma_{r\uparrow})
        \Theta(E_Z,\Omega_R, \omega_{rf}, T_\perp, \Gamma_{\alpha\sigma}),\\
I_{r}^s  & =&e\Omega_R^2(2\Gamma_{r\uparrow}\Gamma_{r\downarrow}
        +\Gamma_{r\uparrow}\Gamma_{l\downarrow}+\Gamma_{r\downarrow}\Gamma_{l\uparrow})
        \Theta(E_Z,\Omega_R, \omega_{rf}, T_\perp, \Gamma_{\alpha\sigma}),
\end{eqnarray}
where the resonance function is
\begin{eqnarray}
\Theta(E_Z,\Omega_R, \omega_{rf}, T_\perp,
\Gamma_{\alpha\sigma})&=&\Upsilon(T_\perp,\Gamma_{\alpha\downarrow})/\Lambda(E_Z,\Omega_R,
\omega_{rf}, T_\perp, \Gamma_{\alpha\sigma}),\\
\Upsilon(T_\perp,\Gamma_{\alpha\downarrow})&=&\Gamma_{l\downarrow}+\Gamma_{r\downarrow}+\frac{2}{T_\perp},
\\
\Lambda(E_Z,\Omega_R, \omega_{rf}, T_\perp,
\Gamma_{\alpha\sigma})&=&4(E_Z-\omega_{rf})^2(\Gamma_{l\uparrow}+\Gamma_{r\uparrow})
(\Gamma_{l\downarrow}+\Gamma_{r\downarrow})
+\Omega_R^2(2\Gamma_{l\uparrow}+ \nonumber \\&&
2\Gamma_{r\uparrow}+\Gamma_{l\downarrow}+\Gamma_{r\downarrow})\Upsilon
+(\Gamma_{l\uparrow}+\Gamma_{r\uparrow})(\Gamma_{l\downarrow}+\Gamma_{r\downarrow})\Upsilon^2.
\end{eqnarray}
We notice that Dong et al.\cite{7} considered a particular case
with all line-width functions being the same, and drew the
conclusions of zero charge current and same values for the spin
current in the two leads. We see from Eq. (7) that the pumped
charge current in the left and right leads has the same magnitude
and direction no matter what  the parameter values. It is a result
of current conservation. Charge current disappears when the
line-width functions satisfy the relation
$\Gamma_{l\uparrow}/\Gamma_{l\downarrow}=\Gamma_{r\uparrow}/\Gamma_{r\downarrow}$.
While the condition to have the same magnitude for the spin
current in the left and right leads is
$\Gamma_{l\uparrow}\Gamma_{l\downarrow}=\Gamma_{r\uparrow}\Gamma_{r\downarrow}$.
In the spin-independent tunneling case
$\Gamma_{\alpha\uparrow}=\Gamma_{\alpha\downarrow}$, charge
current is always zero, and a maximal spin current can be expected
when the time-varying field is resonantly coupled to the dot,
i.e., $\omega_{rf}=E_Z$, and  the dot is coupled to the leads in
an extremely asymmetric way. The ratio $|I_l^s/I_r^s|$ between the
magnitudes of spin current in the left and right leads is directly
proportional to the coupling asymmetry factor $\Gamma_l/\Gamma_r$
in the spin-dependent tunneling case.

  In summary, we have derived explicit expressions for the charge
  and spin current in terms of setup parameters of a quantum dot pump
  device, and discussed a possible way of optimizing the relevant
  parameters to achieve a maximal spin current without charge
  current flow in the leads. The influences of finite coulomb
  interaction and spin-flip process on the charge and spin current in
  such a device in the nonequilibrium situation are expected to be
  more interesting and the study is underway.

\section*{Acknowledgements}

Z.Y.Zeng was supported by the NSFC under Grant No. 10404010, the
Project-sponsored by SRF for ROCS, SEM. This work is supported in
part by a FRG grant of NUS and the DSTA under Project Agreement
POD0410553.

\end{document}